\documentclass[10pt,letterpaper,twocolumn]{article} 

\usepackage{ol2}
\usepackage{amsmath}
\usepackage[square,comma,numbers,sort&compress]{natbib} 
\usepackage{graphicx}
\usepackage{hyperref}

\makeatletter

\newcommand{\Rmnum}[1]{\expandafter\@slowromancap\romannumeral #1@}
\newcommand\relphantom[1]{\mathrel{\phantom{#1}}}
\makeatother

\begin{document}
\setlength{\parindent}{0 cm}

\twocolumn[

\title{Interferometric autocorrelation in the ultra-violet utilizing spontaneous parametric down-conversion}

\author{Patrick Michelberger,$^{1,2,3,*}$ Roland Krischek,$^{1,2}$ Witlef Wieczorek,$^{1,2,4}$   	
Akira Ozawa$^{1,5}$ and Harald Weinfurter$^{1,2}$}

\address{
$^1$Max-Planck-Institut f\"ur Quantenoptik, Hans-Kopfermann-Strasse 1, D-85748 Garching, Germany\\
$^2$Fakult\"at f\"ur Physik, Ludwig-Maximilians-Universit\"at, D-80797 M\"unchen, Germany \\
$^3$Currently with the Clarendon Laboratory, University of Oxford, Parks Road, Oxford OX1 3PU, UK\\
$^4$Currently with the Vienna Center for Quantum Science and Technology, Faculty of Physics, University of Vienna, Boltzmanngasse 5, 1090 Vienna, Austria\\
$^5$Currently with the Institute for Solid State Physics, University of Tokyo, 5-1-5 Kashiwanoha, Kashiwa, Chiba 277-8581, Japan\\
$^*$Corresponding author: p.michelberger1@physics.ox.ac.uk
}

\begin{center}
{\footnotesize \textit{ \copyright 2012 Optical Society of America \cite{published:version}. One print or electronic copy may be made for personal use only. Systematic reproduction and distribution, duplication of any material in this paper for a fee or for commercial purposes, or modifications of the content of this paper are prohibited.
}}
\end{center}

\begin{abstract}
\noindent
Autocorrelation is a common method to estimate the duration of ultra-short laser pulses. In the ultra-violet (UV) regime it is increasingly challenging to employ the standard process of second harmonic generation, most prominently due to absorption in nonlinear crystals at very short wavelengths. Here we show how to utilize spontaneous parametric down-conversion (SPDC) to generate an autocorrelation signal for UV pulses in the infrared. Our method utilizes the $n^{\text{th}}$ order emission of the SPDC process, which occurs for low pumping powers proportional to the $n^{\text{th}}$ power of the UV intensity.  Thus, counting $2n$ down-converted photons directly yields the $n^{\text{th}}$-order autocorrelation. The method, now with detection of near infrared photons with high efficiency, is applied here to the first direct measurement of ultra-short UV-pulses  (approximately $176\, \text{fs}$, center wavelength $390\, \text{nm}$) inside a UV enhancement cavity.
\end{abstract}


] 

Ever since the advent of ultra-short laser pulses, means for determining their duration have been required. 
Although more elaborate techniques exist, interferometric autocorrelation is still the method of choice if only knowledge about the duration of potentially dispersion-broadened pulses is desired \citep{Trebino1,Walmsley}. This method is well established for the visible and infrared regime, yet, it is highly challenging to apply it in the ultra-violet (UV). This is due to high absorption and low detection efficiency of light generated by second-harmonic generation (SHG), the commonly applied non-linear process for autocorrelation measurements. 
Here we demonstrate a method which circumvents these problems by employing spontaneous parametric down-conversion (SPDC) for generating an autocorrelation signal in the infrared. We show its usability by determining the duration of ultra-short UV pulses inside an enhancement cavity, which plays an important role in novel quantum information applications for increasing the rate of down-converted photons \cite{Krischek}. To our knowledge, it is the only method developed so far capable of directly measuring an intra-cavity UV pulse duration.
We will discuss the particular requirements of the new method, explain how to obtain the autocorrelation signals and present our results.

Let us start by relating SPDC-detection to autocorrelation functions. Consider collinear type-\Rmnum{2} SPDC in the single mode approximation \cite{Kwiat}. For small pumping powers \cite{LowPumpingPowers} this yields the state

\begin{equation}
|\Psi \rangle = \sum_{n} \frac{\left( \alpha \cdot E_\mathrm{p} \right)^n}{n!}\left( \hat{a}^{\dagger}_\mathrm{H} \hat{a}^{\dagger}_\mathrm{V}\right)^n |0 \rangle,
\label{equation1}
\end{equation}

with $\hat{a}^{\dagger}_\mathrm{H}$ and $\hat{a}^{\dagger}_\mathrm{V}$ as the creation operators generating a photon from the vacuum $|0 \rangle$ with horizontal (H) or vertical (V) polarization, respectively. The constant $\alpha$ contains the nonlinear crystal's parameters and $E_\mathrm{p}$ represents the pump pulse electric field. Consequently the SPDC process generates in its $n^{\text{th}}$ order emission $2n$ photons ($n$ horizontally and $n$ vertically polarized). Upon coincidence detection of exactly $2n$ SPDC photons one isolates the $n^{\text{th}}$ term of eq. \ref{equation1} and obtains count rates proportional to the pump field $| \left(E_\mathrm{p} \right)^n|^2$. 
The $n^{\text{th}}$-order correlation function \cite{Sala} 

\begin{equation}
g^n(\tau) = \frac{\int |\left( E_\mathrm{p}(t) + E_\mathrm{p}(t-\tau) \right)^n|^2 dt}{\int |E_\mathrm{p}(t)^n|^2 dt + \int |E_\mathrm{p}(t-\tau)^n|^2 dt},
\label{equation4}
\end{equation}

can thus be determined directly from the coincidence detection of $2n$ SPDC photons with a pump pulse timing analogue to the one used in SHG-based autocorrelation: For determining the numerator in eq. \ref{equation4} the SPDC has to be pumped by a coherent superposition of two pulses $E_\mathrm{p}(t)$ and $E_\mathrm{p}(t-\tau)$, delayed by a time $\tau$, whereas for the denominator terms the SPDC pump consists of only one field $E_\mathrm{p}(t)$ or $E_\mathrm{p}(t-\tau)$ at a time. 

\begin{figure}[htb]
\centering
\includegraphics[width=8cm]{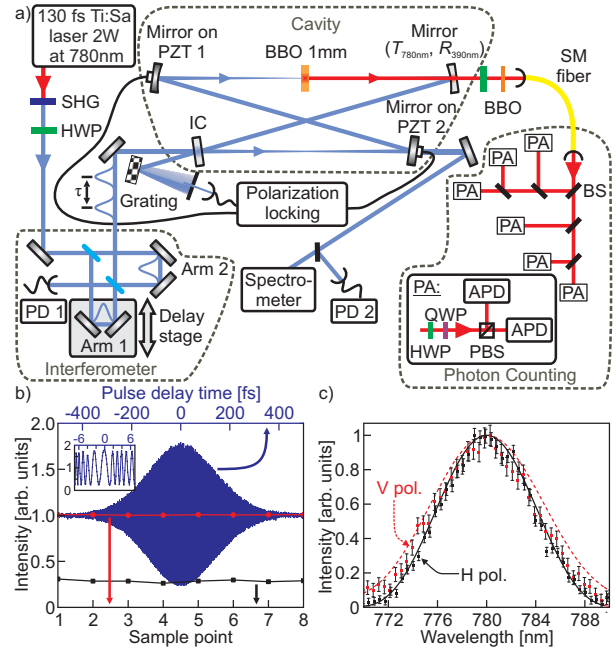}
\caption{\label{figure1} (a) Experimental set-up described in the main text. (b) Interferometric stability demonstration for the set-up; the black and red data points show the intensity on PD $1$ during a measurement run. Red circles correspond to a pulse separation of $\tau = -1850 \, \text{fs}$ and black boxes to $\tau = 0 \, \text{fs}$. The blue line represents the $g^1_{\text{PD1}}(\tau)$ interference pattern on PD $1$ over the entire pulse delay range $\tau$ (visibility $0.77$) and the inset shows a cut-out thereof. (c) Calculated and directly measured spectra of the SPDC photons. Black boxes are the measured signal spectrum (H pol.), red circles display the idler spectrum (V pol.). Appropriately colored lines (straight and dashed, respectively) are the expectations for both.}
\end{figure}

As a first demonstration, we apply this method to determine the temporal duration of UV pulses centred at $390 \, \text{nm}$ circulating in an ultrafast enhancement cavity (Fig. \ref{figure1}(a)). 
Previously this cavity has been shown to enhance the yield of multi-photon entangled states significantly \cite{Krischek}. However, the question arises, how short the pulses still are, given the dispersion of the mirrors and the nonlinear crystal. To answer this, pump pulses, obtained from a titanium sapphire laser with subsequent frequency doubling, are split into two copies by a Michelson interferometer prior to cavity coupling. One output of the interferometer is directly observed by a photodiode (PD $1$), the power level (PD $2$) and the UV spectrum inside the cavity are both measured in transmission behind a cavity mirror. The $n^\text{th}$ order autocorrelation function can now be observed by placing a BBO crystal ($1 \, \text{mm}$ tick, oriented to phasematch type-\Rmnum{2} collinear degenerate SPDC) in one of the foci of the cavity. The emitted photons are coupled out of the cavity by a dichroic mirror and propagated through a single mode (SM) fiber into a linear optics set-up for photon number counting \cite{Wieczorek}. There, photons are split into six distinct spatial modes, each equipped with a polarization analysis unit, enabling coincidence detection between up to six photons simultaneously generated by SPDC.

\begin{figure}[htb]
\centering
\includegraphics[width=8cm]{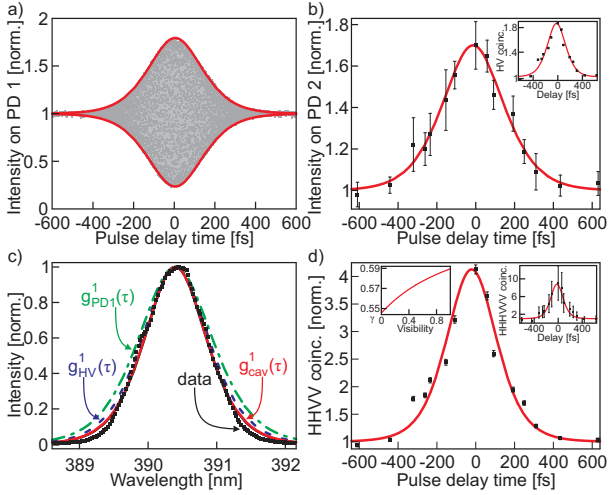}
\caption{\label{figure2} (a) $g^1_{\text{PD1}}(\tau)$-function. Data measured on PD 1 (gray) and the fit according to eq. \eqref{equation2} to the interference extrema (solid red lines). (b) $g^1_{\text{cav}}(\tau)$ interference maxima of the cavity level (black), fitted by eq. \eqref{equation2} (red line). The inset shows $g^1_{\text{HV}}(\tau)$ for the HV coincidences with error bars from Poissonian counting statistics (equal color coding). (c) Intra-cavity UV pulse spectra (black boxes) and a fit to these (solid black line). Also shown are the spectra obtained by FT of $g^1_{\text{cav}}(\tau)$ (solid red line), $g^1_{\text{HV}}(\tau)$ (dashed blue line) and $g^1_{\text{PD1}}(\tau)$ (dotted-dashed green line). (d) $g^2(\tau)$ interference maxima of the HHVV coincidences fitted by eq. \eqref{equation3}. The left inset shows the conversion factor $\gamma$, relating the FWHM of $g^2(\tau)$ to the FWHM pulse duration $\tau_{\text{pulse}}$, over the interference visibility for sech-pulses. In the right inset the HHHVVV coincidences are displayed with the $g^3(\tau)$ expected from the $g^2(\tau)$ parameters. Error-bars and color coding is as in (b).}
\end{figure}

Measuring the $g^n(\tau)$-functions necessitates interferometric stability of the pulse separation $\tau$, phasematching of the nonlinear process over the spectrum of the UV pump and the absence of background noise in detection. 
The latter requirement is satisfied by the low dark count probability of coincidences detected with silicon avalanche single photon detectors. Interferometric stability during the measurement time of $8 \, \text{s}$ per datapoint in the correlation functions is certified by a stable intensity on PD $1$ (Fig. \ref{figure1}(b)). Its fluctuations are negligible compared to the total intensity variations in the $g^1_{\text{PD1}}(\tau)$ interferogram.
It is crucial for this method that the bandwidth of the detected SPDC photons is sufficient for determining the pulse duration. Thus the phasematching, as well as the coupling into the fiber and detection set-up have to be chosen such that no spectral cut-off is introduced. The spectrum observed behind the linear optics set-up matched the the predicted SPDC spectrum modelled with a sech-shaped pump pulse with $\Delta\lambda_{\text{FWHM}}=1.06 \, \text{nm}$ (see discussion of Fig. \ref{figure2}(c) below, measured value $1.06 \pm 0.01 \, \text{nm}$) \cite{Pavel}, as shown in Fig. \ref{figure1}(c).
Additionally it is necessary that the coupling between the spectral cavity modes and the frequency comb modes of the external pulses is constant throughout the measurement \cite{Jones}. 
Since the spectrum of the pulses at the interferometer output is dependent on $\tau$, the locking parameters of the cavity \cite{Krischek} are chosen appropriately to guarantee this \cite{Diplomarbeit}.
 
The first order correlation function is accessible at three points: outside the cavity on PD $1$, inside the cavity on PD $2$ and by the two-photon count-rates (HV) measured in our linear optics set-up (Fig. \ref{figure2}(a) and (b), respectively). Higher order correlations $g^2(\tau)$ and $g^3(\tau)$ (Fig. \ref{figure2}(d)) were measured by counting four- (HHVV) and six-fold coincidences (HHHVVV). 
The phase of the interferometer introducing the time delay $\tau$ was set to multiples of $2\pi$ to maximize the pump power inside the cavity. Thus only the upper envelope in the $g^n(\tau)$-functions has been observed. The interferometer also introduces some residual transverse mode mismatch, resulting in a degraded interference visibility. This leads to lower peak-to-background ratios in the $g^n(\tau)$-functions, yet without loss of phase information or disturbance of the envelope shape. 

For data evaluation, we assume a sech-dependence of the electric fields $E_\mathrm{p}(t) = a \cdot E_0 \cdot \text{sech}\left(\frac{t}{\Delta t} \right)$ and $E_\mathrm{p}(t-\tau) = b \cdot E_0 \cdot \text{sech} \left(\frac{t-\tau}{\Delta t} \right)$ with  $a$ and $b$ used to model reduced interference between the delayed pulse components. We obtain for the envelope of the correlation function \citep{Diplomarbeit,Diels}

\begin{equation}
g^1(\tau) = 1+\frac{2ab}{a^2+b^2} \frac{\tau}{\Delta t \sinh{\left( \tau/\Delta t \right)}}
\label{equation2}
\end{equation}
\begin{align}
g^2(\tau) &= 1+ \frac{18 a^2 b^2}{a^4+b^4} \frac{\left(\frac{\tau}{\Delta t}\right)\cosh{\left(\frac{\tau}{\Delta t}\right)} - \sinh{\left(\frac{\tau}{\Delta t}\right)}}{\sinh^3{\left(\frac{\tau}{\Delta t}\right)}} \nonumber \\  
 					&\relphantom{=} +  \frac{3\left(ab^3+a^3 b \right)}{a^4+b^4}\frac{\sinh{\left(\frac{2 \tau}{\Delta t}\right)}- \left(\frac{2\tau}{\Delta t}\right)}{\sinh^3{\left(\frac{\tau}{\Delta t}\right)}}.
\label{equation3}
\end{align}

Note that the cavity coupling has been optimized for the $E_\mathrm{p}(t)$ field, thus $a$ is set to $1$ and $b$ consequently accounts for the spatial mode mismatch and a reduced cavity coupling of $E_\mathrm{p}(t-\tau)$.
Contrary, for $g^1_{\text{PD1}}(\tau)$ measured on PD $1$ both pulses contribute fully to the background signal and $(2 a b)/(a^2+b^2)$ corresponds to the interference visibility \cite{Diplomarbeit}. The measured data for $g^1_{\text{PD1}}(\tau)$ is shown in Fig. \ref{figure2}(a). The intra-cavity UV power level, yielding $g^1_{\text{cav}}(\tau)$, is given in Fig. \ref{figure2}(b) and $g^1_{\text{HV}}(\tau)$ as deduced from the HV coincidences in the inset. Fits according to eq. \eqref{equation2} are displayed by red lines. We achieve comparable interference visibilities \cite{Diplomarbeit} of $0.71$  inside and $0.77$ outside the cavity. The minimal, i.e., Fourier transform (FT) limited pulse durations $\tau_{\text{FT}}$ can be extracted from the full width half maximum (FWHM)  $\Delta \tau_{g^1}$ of the first order correlation functions by the relation $\tau_{\text{FT}}=0.4048 \cdot \Delta \tau_{g^1}$ for sech-shaped pulses \cite{Diels}. The UV pulses prior to cavity coupling are consequently found to have a minimal duration of $\tau_{\text{FT}}^{\text{ext}}=126 \pm 11 \, \text{fs}$. 
Compared to this we find $\tau_{\text{FT}}^{\text{cav}}= 150 \pm 16 \, \text{fs}$ and $\tau_{\text{FT}}^{\text{HV}}= 140 \pm 17 \, \text{fs}$, respectively, for pulses circulating inside the cavity. This intra-cavity pulse broadening stems from the dispersion of the BBO and air inside the resonator, which reduces the spectral width of light resonant with the cavity \cite{Jones}. The resulting spectral narrowing can also be investigated by examination of the pulse spectrum $S(\lambda)$, derived from $g^1(\tau)$ by FT \cite{Trebino}. These spectra are depicted in Fig. \ref{figure2}(c) together with the directly measured data (fit by $S(\lambda)=\text{sech}^2\left(\Delta t \pi^2 c/ \lambda \right)$ for a sech-pulse). We obtain FWHM spectral widths of $\Delta \lambda_{\text{ext}}= 1.26 \pm 0.01 \, \text{nm}$, $\Delta \lambda_{\text{cav}}= 1.07 \pm 0.04 \, \text{nm}$ and $\Delta \lambda_{\text{HV}}= 1.14 \pm 0.05 \, \text{nm}$. The direct measurement yielded a FWHM $\Delta \lambda_{\text{meas.}}= 1.06 \pm 0.01 \, \text{nm}$ in good agreement with the value of the intra-cavity $g^1(\tau)$-function $\Delta \lambda_{\text{cav}}$.

Next, the intra-cavity pulse duration is determined from HHVV coincidences, yielding $g^2(\tau)$. The recorded data are displayed in Fig. \ref{figure2}(d) together with a fit by eq. \eqref{equation3}. The peak-to-background ratio is $4.18:1$. Ideally $8:1$ is expected, but this ratio is reduced by imperfect mode matching between the interfering pump pulses $E(t)$ and $E(t-\tau)$. Importantly, the conversion factor $\gamma$ between the FWHM $\Delta \tau_{g^2}$ of the $g^2(\tau)$-function and the true pulse duration $\tau_{\text{pulse}} = \gamma \cdot \Delta \tau_{g^2}$ depends on the interference visibility, shown in Fig. \ref{figure2}(d). The value of $0.5895$ \cite{HandbookOfLasers} for perfect interference modifies to $0.582$ for a visibility of $0.75$ extracted from the HHVV coincidence data \cite{Diplomarbeit}. Thus we obtain $\tau_{\text{pulse}}= 176 \pm 14 \, \text{fs}$ for the intra-cavity UV pump pulses, revealing an additional dispersive broadening of approximately $30 \, \text{fs}$ inside the resonator.
Notably, six-photon events, providing additional information about the pulse symmetry \cite{Trebino}, have also been observed (Fig. \ref{figure2}(d)). 
The data agree with the theoretical model, yet, the statistics are not sufficient to allow statements on the pulse shape. An improvement could be obtained here by active interferometer stabilization, allowing longer measurement times.

In summary we have shown a novel method to determine the duration of ultra-short UV pulses. Using SPDC allows one to obtain higher order correlation functions by detection of near-IR photons. The higher order emissions from SPDC enable registration of several correlation functions $g^n(\tau)$ simultaneously. We have applied this method to measure the duration of pulses circulating in an optical enhancement cavity. We estimate pulses to be about $176 \, \text{fs}$ long and dispersion broadened by approximately $30 \, \text{fs}$. The method demonstrated here is a valuable tool for pulse metrology in the UV as it allows efficient analysis also for this wavelength regime. Additionally, it could be extended to achieve a more complete pulse reconstruction, e.g. by frequency resolved optical gating \cite{Trebino1}. 

This work was supported by the EU program Q-ESSENCE (Contract No. 248095), the DFG-Cluster of Excellence MAP, the EU project QAP and the DAAD/MNISW exchange program. P. M. and W. W. acknowledge support by the EU project FASTQUAST and QCCC of the Elite Network of Bavaria, respectively.


\begin{thebibliography}{99}

\bibitem{published:version} The published version of this document is available under: \\ {\footnotesize{\url{https://www.osapublishing.org/ol/abstract.cfm?uri=ol-37-7-1223}}}

\bibitem{Trebino1} D. J. Kane and R. Trebino, IEEE J. Quantum Electron. {\bf 29}, 2 (1993).

\bibitem{Walmsley} C. Iaconis and I. A. Walmsley Opt. Lett. {\bf 23}, 10 (1998).

\bibitem{Krischek} R. Krischek, W. Wieczorek, A. Ozawa, N. Kiesel, P. Michelberger, T. Udem and H. Weinfurter, Nature  Photonics {\bf 4}, 170-173 (2010).

\bibitem{Kwiat} P. G. Kwiat, K. Mattle, H. Weinfurter and A. Zeilinger, Phys. Rev. Lett. {\bf 75}, 24 (1995).

\bibitem{LowPumpingPowers} In our case this is satisfied as powers $P_{\text{UV}} \in \left[ 1.97 \, \text{W}, 3.47 \, \text{W} \right]$ have been used.

\bibitem{Sala} K. L. Sala, G. A. Kenney-Wallace and G. E. Hall, IEEE J. Quantum Electron. {\bf 16}, 9 (1980).

\bibitem{Wieczorek} W. Wieczorek, R. Krischek, N. Kiesel, P. Michelberger, G. T$\acute{\text{o}}$th and H. Weinfurter, Phys. Rev. Lett. {\bf 103}, 020504 (2009).

\bibitem{Pavel} P. Trojek, \textit{Efficient Generation of Photonic Entanglement and Multiparty Quantum Communication}, PhD thesis, LMU Munich, 2007.

\bibitem{Jones} R. J. Jones and J. Ye, Opt. Lett. {\bf 27}, 20, 1848 (2002).

\bibitem{Diplomarbeit} P. Michelberger, \textit{Femtosecond pulsed enhancement cavity for multi-photon entanglement studies}, Diploma thesis, TU Munich, 2009.

\bibitem{Diels} J.-C. M. Diels, J. J. Fontaine, I. C. McMichael and F. Simoni, Appl. Opt. {\bf 24}, 9 (1985).

\bibitem{HandbookOfLasers} F. Tr\"ager \textit{Handbook of Lasers and Optics} (Springer Verlag, 2007)

\bibitem{Trebino} R. Trebino, \textit{Frequency resolved optical gating: The measurement of ultrashort laser pulses} (Kluwer Academic Publishers, 2000)

\end{thebibliography}
\end{document}